# Complex-dynamic cosmology and emergent world structure[*]

ANDREI P. KIRILYUK[†]

Institute of Metal Physics
36 Vernadsky Bd., Kiev-142, Ukraine 03142

Universe structure emerges in the unreduced, complex-dynamic interaction process with the simplest initial configuration (two attracting homogeneous fields). The unreduced interaction analysis gives intrinsically creative cosmology, describing the real, explicitly emerging world structure with dynamic randomness on each scale. Without imposing any postulates or entities, we obtain physically real space, time, elementary particles with their detailed structure and intrinsic properties, causally complete and unified version of quantum and relativistic behaviour, the origin and number of naturally unified fundamental forces, and classical behaviour emergence in closed systems. Main problems of standard cosmology and astrophysics are consistently solved in this extended picture (without introduction of any additional entities), including those of quantum cosmology and gravity, entropy growth and time, "hierarchy" of elementary particles, "anthropic" difficulties, space-time flatness, and "missing" ("dark") mass and energy. The observed universe structure and laws can be presented as manifestations of the universal symmetry (conservation) of complexity providing the unified, irregular, but exact (never "broken") Order of the World.

## 1. Standard vs complex-dynamic cosmology

Contrary to experimental, observational successes in modern astro-physics, the explanatory power of the corresponding cosmological theories remains limited, so that the number of unsolved problems only grows, while those considered to be "solved" often resemble rather a "plausibly" looking adjustment of artificially introduced, abstract entities and free parameters (see e.g. ref. [1] for a critical review). Without going into detailed discussion of those difficulties, we only note here that one can see a possible general origin of such situation, which is inherent in the conventional theory, but have specific, more distinct manifestations in cosmology. As it has first been emphasized by Bergson [2] and confirmed by further science development (see e.g. [3]), conventional science methods do *not* describe the explicit structure emergence as such, but are limited instead to postulation of already existing structure configuration, properties, and formal, imitative "evolution" (in the form of empirically guessed "laws", "principles", "models", "equations", etc.). Whereas such approach can be relatively useful in the study of simple, easily measurable and "smoothly" evolving objects (the canonical case of "Newtonian mechanics"), it will be much less efficient in explanation of the origin and dynamics of systems, such as the universe, that cannot be simply "postulated" with all their observed properties because they undergo strong, qualitative changes of configuration (explicit emergence of structure) involving many diverse, hierarchically organised, and entangled elements.

In other words, the *true* cosmology should describe the unreduced, *explicit* formation of a complicated structure, which just remains obscure in the conventional theory framework. A related difficulty of the latter is that it *cannot* consistently *solve* any realistic, many-body *interaction problem*, always resorting to one or another simplified "model" or "perturbative" approximation, whereas it is clear that such unreduced interaction process underlies any real structure formation. In particular, the standard theory cannot provide the unambiguous, universal origin of the major *property of mass (and energy)*, operating instead with its measurable inertial and gravitational manifestations. Although this problem could remain among "less important" ones in "Newtonian" science fields, the difficulties with "dark" mass and energy "suddenly" emerge now on the global scale as quite important, if not fatal, defects of the *whole* world picture.

---

[*]Report presented at the International Workshop on Frontiers of Particle Astrophysics (Kiev, June 21-24, 2004), http://www.physics.ucla.edu/~kusenko/Kiev/.

[†]Address for correspondence: Post Box 115, Kiev-30, Ukraine 01030. E-mail address: kiril@metfiz.freenet.kiev.ua.

In this report we describe a *new, qualitatively extended cosmology framework*, based on the *unreduced*, truly "exact" *solution of arbitrary interaction problem* that gives *explicit* emergence of *real* structures without any artificial simplification and leads to the rigorously derived, truly *universal* concept of *dynamic complexity* [4-17]. This unreduced dynamic complexity is quite different from the existing mechanistic imitations of "complexity" in the conventional theory and *unifies* the *qualitatively extended* versions of dynamical chaos, self-organisation, self-organised criticality, fractality, adaptability, etc.

We start with showing how the fundamental entities and properties of the universe, including physically specified space and time, elementary particles, their properties, interactions and dynamics, *explicitly emerge* in the *provably simplest* initial configuration of interaction process, comprising two structureless, omnipresent, physically real fields, homogeneously attracted to each other (section 2). It is important that we obtain *together* the *main entities* (space, time, particles), their *properties* (space discreteness and number of dimensions, irreversible time flow, mass-energy, charge, spin, interactions), and *dynamical laws* (quantum and relativistic behaviour) within the *same, intrinsically unified* concept of complexity, using a rigorous derivation procedure and *no* additional, postulated laws or entities besides the evidently "minimal" starting interaction configuration (section 3). We show then how the naturally emerging, truly *dynamic* properties of *complexity* and *chaoticity* in a system with interaction give rise to all higher-level structures and solve the difficulties of the conventional theory which neglects those properties because of its artificial reduction and therefore loses inevitably the essence of such major properties as mass and energy (sections 4, 5).

We emphasize the *intrinsically unified* and *reality-based* character of the proposed solution to all the main, quite diverse problems of the scholar theory, consistently derived simply by the *unreduced, universally nonperturbative solution* of an *arbitrary (generic) interaction problem*, which confirms the power of unreduced scientific logic and reveals the genuine, *exact* origin of the standard theory limitations and difficulties as being due to the dynamically single-valued, zero-complexity approximation of the canonical approach that neglects all emerging system realisations but a single, "averaged" one. The ultimate, mathematically exact expression of the obtained unification is provided by the *universal symmetry, or conservation, of complexity* that determines the emergence of all structures of the universe and laws of their behaviour and therefore constitutes the genuine, unique Order of the World (section 2) [4-6].

## 2. Universe structure emergence in the unreduced interaction process development

No structure can emerge without interaction. A universe structure formation should start from the simplest possible (least structured) interaction configuration, which is still able to produce explicitly the observed real structures. The most structureless configuration of a physically real system with interaction is given by two homogeneous (*effectively structureless*), uniformly interacting entities represented by two *physically real* fields, or *protofields*, which are attracted to each other and whose composition (of sufficiently small elements) does not play any *essential, direct* role in the following structure formation [4,5,11-17]. Interaction between protofields supposes their different physical qualities designated as *gravitational* protofield (or medium) and *electromagnetic* (e/m) protofield, since we show later that they are responsible for the *emerging* (and universally present) gravitational and e/m interactions, respectively. The structureless protofields are omnipresent by definition and therefore *cannot* be related to any postulated (let alone "hidden") spatial "dimensions", time "variables", mathematical structures and laws, etc., none of which may have a sense at this initial stage (cf. recent imitations of this approach within so-called "brane-world" scenarios of the unitary theory [18-20]). The extended, complex-dynamical and physically real versions of those entities and laws are *consistently derived* in our theory starting from the *existence equation* that describes the above simplest protofield interaction without any limitation or model assumption [4-6,11-17]:

$$\left[ h_\text{g}(\xi) + V(\xi,q) + h_\text{e}(q) \right] \Psi(\xi,q) = E \Psi(\xi,q), \tag{1}$$

where $h_\text{g}(\xi)$ and $h_\text{e}(q)$ are "generalised Hamiltonians", representing the internal dynamical properties of the free (non-interacting) gravitational and e/m protofields in terms of a measure of the unreduced dynamic complexity defined below, $V(\xi,q)$ is the corresponding measure of (generally arbitrary) potential of attractive interaction between protofields, whose physically different degrees of freedom are represented by $\xi$ (gravitational medium) and $q$ (e/m protofield), $\Psi(\xi,q)$ is the compound system *state-function* characterising completely its configuration and properties, and $E$ is the eigenvalue of the generalised Hamiltonian for the compound system. Note that eq. (1), as well as its further analysis, does not assume anything beyond the initial system configuration and can eventually take the form of various "model", including "nonlinear", equations (although we show below, in a self-consistent way, that its "Hamiltonian" form is indeed absolutely universal [4-6,11-17]).

It is convenient to express the same problem in terms of e/m protofield excitations (local deformations):

$$h_e(q)\varphi_n(q) = \varepsilon_n \varphi_n(q), \tag{2a}$$

$$\Psi(\xi,q) = \sum_n \psi_n(\xi)\varphi_n(q), \tag{2b}$$

where $\{\varphi_n(q), \varepsilon_n\}$ is the complete set of orthonormal eigen-solutions of the free e/m protofield Hamiltonian $h_e(q)$. Substituting eq. (2b) into eq. (1) and using the standard procedure of scalar-product separation (e.g. by integration), we obtain the equivalent system of equations for $\{\psi_n(\xi)\}$:

$$\left[h_g(\xi) + V_{nn}(\xi)\right]\psi_n(\xi) + \sum_{n' \neq n} V_{nn'}(\xi)\psi_{n'}(\xi) = \eta_n \psi_n(\xi), \tag{3}$$

where $\eta_n = E - \varepsilon_n$ and

$$V_{nn'}(\xi) = \int_{\Omega_q} dq\, \phi_n^*(q) V_{eg}(q,\xi) \phi_{n'}(q). \tag{4}$$

Note that eqs. (3) express the same problem configuration as eq. (1), but now in terms of the "physically specified" degrees of freedom of the e/m protofield (considered to be "known"), which should be possible for any correct model of the protofield dynamics.

A usual perturbative (or model) analysis of system (3) would reduce it to approximate but separated, "integrable" equations of the form

$$\left[h_g(\xi) + V_{nn}(\xi) + \tilde{V}_n(\xi)\right]\psi_n(\xi) = \eta_n \psi_n(\xi), \tag{5}$$

where an integrable potential $\tilde{V}_n(\xi)$ of a "mean-field" approximation can vary between zero and an extreme configuration, such as

$$\tilde{V}_n(\xi) = \sum_{n'} V_{nn'}(\xi). \tag{6}$$

If, however, we avoid any perturbative reduction of system (3) and try to find its unreduced solution by the method of substitution, using the standard Green function technique, then we arrive at problem formulation in terms of generalised *optical, or effective, potential (EP)* [4-17,21,22]. The latter is a well-known method, but is used routinely in its reduced, perturbative approximation (see e.g. [21]). Direct analysis of the unreduced EP expression shows that the original problem has the *redundant* number of locally "complete" and thus *mutually incompatible* (but *equally real*) solutions called system and problem *realisations* [4-17,22]. Therefore the truly *complete general solution* to a problem is given, in terms of system "density" $\rho(\xi,q)$ (generalising also other measured quantities), by the *causally probabilistic* sum over redundant realisations, which *permanently replace one another* in a *dynamically random* order thus defined:

$$\rho(\xi,q) \equiv |\Psi(\xi,q)|^2 = \sum_{r=1}^{N_\Re} {}^\oplus \rho_r(\xi,q), \tag{7}$$

where $N_\Re$ is the total number of realisations (it's maximum value is equal to the number $N_\xi$ of degrees of freedom, or local dynamical modes, of the gravitational protofield, involved in the interaction process [4-17]), $\rho_r(\xi,q) = |\Psi_r(\xi,q)|^2$ is the generalised density of the *r*-th realisation with the state-function $\Psi_r(\xi,q)$, and the sign $\oplus$ designates the special, dynamically probabilistic meaning of the sum, outlined above.

The system state-function $\Psi_r(\xi,q)$ entering the general solution, eq. (7), is obtained in the unreduced EP method in the form [4-17]:

$$\Psi_r(\xi,q) = \sum_i c_i^r \left[ \varphi_0(q)\psi_{0i}^r(\xi) + \sum_{n,i'} \frac{\varphi_n(q)\psi_{ni'}^0(\xi) \int_{\Omega_\xi} d\xi' \psi_{ni'}^{0*}(\xi') V_{n0}(\xi') \psi_{0i}^r(\xi')}{\eta_i^r - \eta_{ni'}^0 - \varepsilon_{n0}} \right], \quad (8)$$

where $\varepsilon_{n0} \equiv \varepsilon_n - \varepsilon_0$, $\{\psi_{0i}^r(\xi), \eta_i^r\}$ are *r*-th realisation eigen-solutions (eigenfunctions and eigenvalues) of the *effective* existence equation (obtained from equation for $\psi_0(\xi)$ in the system (3)):

$$h_g(\xi)\psi_0(\xi) + V_{\text{eff}}(\xi;\eta)\psi_0(\xi) = \eta\psi_0(\xi), \quad (9)$$

the EP operator $V_{\text{eff}}(\xi;\eta)$ is defined by its action:

$$V_{\text{eff}}(\xi;\eta_i^r)\psi_{0i}^r(\xi) = V_{00}(\xi)\psi_{0i}^r(\xi) + \sum_{n,i'} \frac{V_{0n}(\xi)\psi_{ni'}^0(\xi) \int_{\Omega_\xi} d\xi' \psi_{ni'}^{0*}(\xi') V_{n0}(\xi') \psi_{0i}^r(\xi')}{\eta_i^r - \eta_{ni'}^0 - \varepsilon_{n0}}, \quad (10)$$

and $\{\psi_{ni}^0(\xi), \eta_{ni}^0\}$ are eigen-solutions of a truncated system of equations:

$$[h_g(\xi) + V_{nn}(\xi)]\psi_n(\xi) + \sum_{n' \neq n} V_{nn'}(\xi)\psi_{n'}(\xi) = \eta_n \psi_n(\xi) \quad (11)$$

($n, n' \neq 0$ in eqs. (8)-(11) and everywhere below, contrary to the starting system of equations (3) that includes an equation for $\psi_0(\xi)$).

Plurality of locally complete (usual) solutions of eq. (9), or *dynamic multivaluedness* of the unreduced problem, which gives rise to the major property of *causal randomness*, eq. (7), follows from the self-consistent, *dynamically* emerging, or *essentially nonlinear*, dependence of the unreduced EP, eq. (10), on the eigen-solutions to be found [4-17]. We thus obtain also the *dynamically derived, a priori probability* $\alpha_r$ of each *r*-th realisation emergence:

$$\alpha_r = \frac{1}{N_\Re} \quad (r = 1, ..., N_\Re), \quad \sum_{r=1}^{N_\Re} \alpha_r = 1. \quad (12a)$$

In the general case, at a higher level of dynamics, we shall have

$$\alpha_r = \frac{N_r}{N_\Re} \left( N_r = 1, ..., N_\Re; \sum_r N_r = N_\Re \right), \quad \sum_r \alpha_r = 1, \quad (12b)$$

where $N_r$ is the number of "elementary realisations" obtained above and entering the *r*-th actually observed, compound realisation. Note that the conventional, perturbative models of eqs. (5)-(6) correspond to rejection of all system realisations but a single, "averaged" one. We call this property of usual, "exact" solutions *dynamic single-valuedness* and the whole standard-theory reduction *dynamically single-valued, or unitary, solution* and approach.

Another major property of the unreduced solution, closely related to dynamic multivaluedness, is *dynamic entanglement* of the inter-acting system components (protofields in this case), expressed by the dynamically weighted products of different component eigenfunctions depending on their respective "degrees of freedom" ($\xi, q$) in the unreduced state-function expression, eq. (8). Dynamic entanglement provides the physical meaning of *interaction* as such, as well as the *rigorous* expression of tangible *quality* of interaction products, absent in any unitary theory describing only an abstract, external "envelope" of a real structure. The property of dynamic entanglement is further amplified by that of *dynamically probabilistic fractality* of the unreduced solution, which extends essentially the ordinary, dynamically single-valued fractality and is obtained by repeated use of the same, universal EP method in order to solve truncated systems of equations, starting from eqs. (11), whose solutions enter the expressions for previous level of structure (see eqs. (8), (10)) [4,8]. One obtains thus the whole hierarchy of not only entangled, but *permanently probabilistically, interactively changing* and thus *dynamically adapting* realisations of the emerging system structure.

It is not difficult to find the emerging realisation configuration for two attracting, initially homogeneous protofields [4,5,10,11,13,17]. The resonant-denominator structure of the state-function and EP expressions, eqs. (8), (10), in combination with "cutting" integrals in the numerators, shows that the magnitude of the state-function components for each particular (*r*-th) realisation is concentrated around certain eigenvalue $\eta_i^r$ for that realisation, which can be conveniently designated as $\eta_r^r$ and interpreted as the centre of the dynamically emerging, local concentration of the attracting protofield density, or *emerging physical space point* and its *coordinate*. This local *dynamical squeeze* of the initially totally homogeneous protofield system appears to be *inevitable* physically, for the real, *unreduced* interaction dynamics: every small density increase of a protofield will provoke a self-amplifying chain of further density increase of both protofields around that location because the larger is the density of one of the protofields, the stronger is attraction between the protofields at that place. That omnipresent *dynamic instability* of the unreduced protofield interaction, accompanied and assisted by the above dynamic entanglement, is absent in any unitary approximation, cutting the emerging interaction links and therefore predicting only small deviations from the initial configuration. In the unreduced analysis it leads to maximum local squeeze, or *dynamic reduction*, of the attracting protofields around a location, or (emerging) physical point, which is chosen among other neighbouring, *equally probable* locations in a *causally (dynamically) random* way, in full agreement with the above rigorously derived expressions for realisation structure and probability, eqs. (8), (10), (12). The maximum squeeze of entangled protofields, determining the fully developed structure of a "regular" system realisation, is limited by finite protofield compressibility, and is naturally followed by the reverse process of protofield *disentanglement* and *extension* to the initial, quasi-homogeneous state, which is initiated and governed by the same system instability as the previous phase of reduction.

One obtains thus the emerging, *physically specified* and totally *real* dynamical *structure of (massive) elementary particle*, such as the electron, in the form of *unceasing* periodic cycles of dynamic reduction and extension of two attracting protofields, where the centre of each next reduction is chosen by the system in a *dynamically (truly) random* fashion. We call this explicitly emerging, internally entangled, permanently changing, and spatially chaotic particle structure *quantum beat process* [4,5,11-17]. Its reality is confirmed by the properties of the unreduced solution within the generalised EP formalism, eqs. (7)-(12). In particular, the latter contains not only the locally squeezed structure of "regular" realisations described above, but also one specific, extended realisation with a "loose", chaotically fluctuating structure that describes the disentangled system state during transition between two squeezed, "regular" realisations. It is this transient state *called intermediate, or "main", realisation* that corresponds to effectively weak interaction value of a perturbative approximation (eqs. (5)-(6)) and constitutes the *physically real* particle *wavefunction*, which represents the *totally causal* extension of the unitary quantum wavefunction (*artificially* mystified because of *dynamically single-valued* "modelling") and due to the unrestricted universality of our analysis can be further extended to *any* kind of system and level of world dynamics (also in the form of generalised *distribution function*) [4-6,12-17]. This physically real, interaction-driven *duality* between squeezed and extended state/phase of the quantum beat process within the elementary particle evokes its another definition as *elementary field-particle* [4,5,11-17].

The emerging *length scale* $\Delta x$ of the quantum beat process is rigorously defined by the distance between neighbouring regular realisations as given by the eigenvalue separation $\Delta_r \eta_i^r$ for different *r*, found from the unreduced EP formalism, eqs. (9)-(10), $\Delta x = \Delta_r \eta_i^r$. It is the length of the elementary, real *quantum jump* of the squeezed, "corpuscular" state of the particle, or *virtual soliton*, between its two regular realisations within the quantum beat process, which is equal to the Compton length $\lambda_C$ for the electron, $\Delta x = \Delta_r \eta_i^r = \lambda_C$ [4,5,11-17]. Another characteristic length scale, determining the size of the virtual soliton, or "particle" (electron) as such, is given by the generic eigenvalue separation $\Delta_i \eta_i^r$ for different $i$, equal to the "classical radius" $r_e$ of the electron, $\Delta_i \eta_i^r = r_e$ (see also section 3.2). We thus obtain the *physically real*, naturally *discrete*, *dynamically entangled*, and *chaotically* changing *space*. Since we have obtained the well-defined *events* of dynamic reduction-extension, we can define the *physically real time*, whose *unceasing flow* is derived as *permanent realisation change* of *dynamically multivalued* protofield interaction process (quantum beat), *intrinsic irreversibility* is provided also by the *dynamically chaotic* sequence of realisations (reduction centres), and *elementary interval* $\Delta t$ can be obtained as $\Delta t = \Delta x/c$, where $\Delta x = \Delta_r \eta_i^r$ is the above space element and $c$ is the speed of perturbation propagation in the e/m protofield interacting with the gravitational protofield (known as the *speed of light*). It is clear that $\Delta t = \tau$, where $\tau$ is the period of quantum beat.

A big number of *different* elementary field-particles will emerge in the described way in the initially homogeneous system of two interacting protofields. This follows from the same basic property of dynamic multivaluedness and its hierarchical fractal structure. Local quantum beat processes can have several major realisations with essentially different EP magnitude, where relatively small amplitudes form (compound) realisation of lighter particles (leptons) with weaker relation to the gravitational protofield, while much larger amplitudes constitute realisation of hadrons with essential entanglement of e/m and gravitational protofields. Each of such "big" compound realisations of the first level of interaction structure can contain various particle subspecies and then end up in splitting into numerous individual particles, situated at different (emerging) locations and represented by a certain level of fractal hierarchy of dynamic multivaluedness, described above as quantum beat process within each (massive) particle. Higher levels of (weaker) interaction between these entities of the first complexity level start then naturally emerge (see below), but the factor of deep *cosmological* importance at this and higher levels of structure emergence is their *intrinsic, dynamic adaptability* determined by the *self-consistent* dependence of the *unreduced*, nonperturbative EP, eqs. (8)-(10), on the emerging structure parameters (exemplified by the eigenvalues $\eta$). Thus, any new particle emergence increases the protofield tension, and when the latter is high enough, no more particles can form (for a given interaction magnitude). Therefore the protofield interaction strength dynamically determines the number (mass density) of particles in the universe. One obtains thus a *self-tuning universe* that avoids, simply due to its unrestricted interaction problem solution, any "anthropic" problems or "catastrophically" adjusted values of universal constants [4,5,11-13,17] (see also sections 3.2, 3.3).

The quantity of *dynamic complexity* as such of *any* real interaction process and emerging structure can now be *universally* defined as a growing function of the total number of its realisations (*explicitly obtained* from the unreduced problem solution) or of their rate of change, equal to zero for the unrealistic case of only one system realisation (it is this extreme simplification of reality that is exclusively considered in the unitary theory, which explains, as we continue to show below, all its "old" and "new" problems).[‡] The physically real, dynamically emerging space and time defined above constitute *two universal manifestations* of the unreduced complexity, characterising already a single realisation structure (space) and change (time). We shall proceed now to the most important *forms* and *measures* of dynamic complexity, representing the totality of system realisations and thus its causally complete structure and dynamics.

A universal measure of complexity is provided by the simplest combination of *independent* space and time elements, known as *action* and acquiring now an extended, *universal* and *essentially nonlinear* meaning, $\Delta \mathcal{A} = -E\Delta t + p\Delta x$, where $\Delta \mathcal{A}$ is the action-complexity increment, while $E$ and $p$ are coefficients identified as energy and momentum. They represent universal *differential* measures of complexity related to the *integral* measure of action:

$$E = -\frac{\Delta \mathcal{A}}{\Delta t}\Big|_{x=\text{const}}, \qquad p = \frac{\Delta \mathcal{A}}{\Delta x}\Big|_{t=\text{const}}. \tag{13}$$

The action-complexity increment $\Delta \mathcal{A}$ for a field-particle at rest corresponds to one quantum beat cycle and explains the meaning of Planck's constant, $\Delta \mathcal{A} = -h$, after which eq. (13) takes the form:

$$E_0 = -\frac{\Delta \mathcal{A}}{\Delta t} = \frac{h}{\tau_0} = h\nu_0 \ , \tag{14a}$$

---

[‡] It is clear that dynamic complexity thus defined is also a measure of dynamical randomness, or chaoticity, or (generalised) entropy (see the end of this section).

where $E_0$ is the particle rest energy, $\tau_0 = \Delta t$ is the quantum beat period at rest, and $\nu_0 = 1/\tau_0$ is its frequency. Since the rest energy results from spatially *chaotic* wandering of the virtual soliton within the particle wave field, it possesses the *causally substantiated* property of *inertia*, as expressed by the *rest mass* $m_0$: $E_0 = m_0 c^2$, where $c^2$ is a coefficient for the moment (rigorously identified later as the square of light velocity). We can understand now the true meaning of a basic relation used by Louis de Broglie for derivation of his formula for the particle wavelength [23,24] as expression of chaotic, essentially nonlinear quantum beat dynamics [4,11-17]:

$$m_0 c^2 = h\nu_0 \ . \tag{14b}$$

The state of rest corresponds to the local minimum of complexity-energy and homogeneous distribution of realisation probabilities. *Motion* is *rigorously defined* as increased complexity and inhomogeneous realisation probability distribution ($p \neq 0$), so that

$$\frac{\Delta \mathcal{A}}{\Delta t} = \frac{\Delta \mathcal{A}}{\Delta t}\Big|_{x=\text{const}} + \frac{\Delta \mathcal{A}}{\Delta x}\Big|_{t=\text{const}} \frac{\Delta x}{\Delta t} ,$$

which transforms eq. (14a) into

$$E = -\frac{\Delta \mathcal{A}}{\Delta t} + \frac{\Delta \mathcal{A}}{\lambda}\frac{\Delta x}{\Delta t} = \frac{h}{T} + \frac{h}{\lambda}\upsilon = hN + p\upsilon , \tag{15}$$

where $E = h/\tau = h\nu$ is the total energy, $\tau \equiv (\Delta t)|_{x=\text{const}}$ is the quantum beat period at a fixed space point, $\nu = 1/\tau$, $\lambda \equiv (\Delta x)|_{t=\text{const}} = \lambda_B = h/p$ is the space element of the moving field-particle, known as de Broglie wavelength $\lambda_B$, $T = \Delta t$ is the "total" quantum beat period ($T \neq \tau$), $N = 1/T$, and $\upsilon = \Delta x/\Delta t$ is the velocity of global field-particle motion. Since the latter emerges only as an *average tendency* in the *chaotic* virtual soliton wandering with the single jump velocity $c$ (the material *speed of light* defined above), one can express the *ensuing* difference between $c$ and the global motion velocity $\upsilon$ by the generalised "relativistic" *dispersion relation* [12,13,17]:

$$p = E\frac{\upsilon}{c^2} = m\upsilon , \tag{16}$$

where the total mass $m \equiv E/c^2$, now by *rigorously obtained* definition. Using eq. (16), one gets the known, but now causally derived, realistically explained expression for the de Broglie wavelength:

$$\lambda = \lambda_B = \frac{h}{m\upsilon} \ . \tag{17}$$

In addition, the dispersion relation thus derived from causal *quantum* dynamics, $p = m\upsilon$, provides (upon time differentiation) the *rigorous* substantiation for Newton's laws of *classical* mechanics (in their relativistic version), thus demonstrating the essential role of the underlying *complex* (multivalued) interaction dynamics also for those higher, classical levels of world dynamics.

Using the relation between *p* and *E* of eq. (16) and the total energy expression through the quantum beat period ($E = h/\tau$) in eq. (17), we get the *rigorously derived* expression of *time relativity* and its causal origin in the underlying complex interaction dynamics:

$$\tau = T\left(1 - \frac{\upsilon^2}{c^2}\right). \tag{18}$$

Time goes more slowly "within" the moving elementary field-particle ($T > \tau$) because the time flow is *produced* by the *same*, essentially nonlinear and dynamically multivalued, quantum beat process that gives rise to global motion. If we use the straightforward relation to the quantum beat period at rest, $T\tau = (\tau_0)^2$ [4,12,13,17], we get the canonical expression of (now causally derived) time relativity:

$$T = \frac{\tau_0}{\sqrt{1 - \frac{\upsilon^2}{c^2}}} \quad \text{or} \quad N = \nu_0\sqrt{1 - \frac{\upsilon^2}{c^2}} \ . \tag{19}$$

Combination of eqs. (15)-(17), (19) provides now the explicit expression of the unified, causal understanding of quantum and relativistic behaviour of a field-particle obtained as the quantum beat process:

$$E = h\nu_0\sqrt{1-\frac{v^2}{c^2}} + \frac{h}{\lambda_\text{B}}v = m_0 c^2\sqrt{1-\frac{v^2}{c^2}} + \frac{m_0 v^2}{\sqrt{1-\frac{v^2}{c^2}}} \quad . \tag{20}$$

The quantum wave equations (of Klein-Gordon, Dirac and Schrödinger) can be *derived* from eq. (20) by *causal quantisation*, expressing multivalued dynamics in terms of intermediate, delocalised realisation of the wavefunction [4,5,12,13,16,17] (see also below).

Elementary field-particles, causally obtained thus as complex-dynamical quantum beat processes, form the entities of the first level of emerging real-world structure, or first *level of complexity*. Due to the *physically* unified world construction of two interacting proto-fields, the entities of the first level start interacting among them and form higher levels of complex-dynamical world structure by the same, universally described development of unreduced interaction process. It is easy to see that the number, physical origin, and properties of the four "fundamental forces" between particles obtain a transparent explanation within this theory [5,12-17], designated as *quantum field mechanics*. Long-range interaction through the e/m and gravitational protofield gives the omnipresent e/m and gravitational interactions, respectively, whereas short-range ("contact") interaction between the protofield elements (poorly resolved as such) appears as "weak" and "strong" interaction forces, where one can clearly see the origin of the (known) unification of e/m and weak interactions (transmitted by the e/m protofield) and similar (but unrecognised) unity between the gravitational and strong interaction. Moreover, *all the four interactions* are naturally, *dynamically unified* within each elementary (hadronic) particle-process, especially in the maximum squeeze state of its unceasing quantum beat pulsation. The physical origin of the gravitational protofield, or medium, can also be causally understood now as a dense, viscous form of "quark matter" (or "condensate"), where the famous "confinement" of quarks acquires an equally transparent origin. Photons, on the other hand, can be interpreted as relatively weak, and therefore quasi-regular and massless, excitations of the e/m protofield, stabilised by attraction to the gravitational protofield (and being thus closer to usual, regular solitons).

One obtains also the dynamic, physical interpretation of *electric charge* (as quantum beat pulsation with a *fixed temporal phase*), its "quantized" value, and two "opposite" types (as quantum beat *synchronisation* in the e/m medium) [4,12-17], where the *quantized* e/m interaction by "exchange of photons" (during the "extended" phase of quantum beat) acquires now a *physically real* sense. The property of *spin* and related *magnetic field* effects obtain a uniquely consistent explanation as highly nonlinear *vortex* dynamics of the reduction-extension process within every quantum beat cycle.

Complex-dynamic interaction development between field-particles leads to causally understood processes of *true quantum chaos* (absence of dissipation) [4,5,9], *quantum measurement* (small dissipation) [4,10], and *classical (permanently localised) behaviour* emergence in an elementary bound, *closed* system (like atom) [4,5,12-17], without any ambiguous "decoherence by environment". Classical behaviour emerges as a next, *superior level of complexity* that gives rise, in its turn, to all higher complexity levels by further development of the same unreduced, intrinsically unified interaction process between two initially homogeneous protofields. The *complete macroscopic world structure and dynamics* is thus *explicitly obtained* from that starting "minimal" interaction configuration, where such persisting "cosmological" problems as origin and emergence of space and time, the "wavefunction of the universe", classicality emergence, and quantum gravity are naturally resolved, together with other ones from particular fields of fundamental physics, within the *intrinsically unified* description of complex protofield interaction dynamics [4-7,12-17].

The unrestricted universality of structure emergence description finds its most complete expression in the *universal symmetry (or conservation) of complexity* [4-6,13,17], which provides the *unified, causally complete* extension of *all* (correct) dynamic equations, laws and principles, remaining unexplained (postulated), separated, and often contradictory within the dynamically single-valued projection of reality in the standard, unitary theory. The causally specified qualitative *change* and *explicit structure emergence* in the universal science of complexity permit us to introduce two major *forms of dynamic complexity*. One of them is called *dynamic information*, *I*, and expresses the real interaction complexity *before* any structure emergence has actually begun. It generalises the usual notion of "potential energy" and is actually given (in its integral version) by the *generalised action*, $\mathcal{A}$, introduced above. The second universal form of complexity is called *dynamic entropy*, $S$, and characterises the dynamic complexity of *already appeared* structures (it generalises the usual notions of "kinetic" and "heat" energy). The symmetry, or conservation, of complexity means that every process occurs so that the sum of dynamic information and entropy, or *total dynamic complexity*, *remains constant*, $C = I + S = \text{const}$, which means that *always decreasing* dynamic information (expressing system "potentiality") is transformed into the dual, *always growing* complexity form of dynamic entropy, $\Delta I = -\Delta S < 0$. The "first" and "second" laws of thermodynamics are thus essentially extended to *any* kind of system or process, *unified* in a single, dynamic *symmetry*, and liberated from unpleasant skewness of the usual second law (which resolves the related cosmology problems, see section 4). Contrary to any unitary symmetry, the symmetry of complexity is always *exact* (never "broken"), but gives more or less *irregular* structures.

Dynamic version of the symmetry of complexity is obtained if we divide its differential expression, $\Delta \mathcal{A} = -\Delta S$ (where $\Delta \mathcal{A} = \Delta I$ and $\Delta S$ are increments of dynamic information and entropy), by a (dynamically) discrete time increment $\Delta t |_{x=\text{const}}$, to get the *generalised Hamilton-Jacobi equation* [4-6]:

$$\frac{\Delta \mathcal{A}}{\Delta t}\Big|_{x=\text{const}} + H\left(x, \frac{\Delta \mathcal{A}}{\Delta x}\Big|_{t=\text{const}}, t\right) = 0 , \qquad (21)$$

where the Hamiltonian $H(x,p,t)$ expresses the entropy-like form of differential complexity, $H = (\Delta S/\Delta t)|_{x=\text{const}}$, and eqs. (13) are taken into account. Because of the dynamically random order of emerging realisations, the dynamic information can only decrease with each real time step, which means that the total time derivative of action, or (generalised) *Lagrangian*, $L = \Delta \mathcal{A}/\Delta t = pv - H$, is always negative:

$$L < 0 \quad \Rightarrow \quad E, H\left(x, \frac{\Delta \mathcal{A}}{\Delta x}\Big|_{t=\text{const}}, t\right) > pv \geq 0 . \qquad (22)$$

We obtain in that way the *rigorously derived* expression of the *arrow of time* always oriented, according to eq. (22), in the direction of growing dynamic entropy. Note that for a system globally at rest ($p = 0$), this condition is equivalent to strict positivity of (generalised) complexity-energy (or Hamiltonian): $E, H > 0$.

The *dynamic quantisation* condition reflects permanent realisation change process through the intermediate state of wavefunction, $\Psi$, and means that this state and the total system complexity remain the same during each cycle of realisation change [4-6,12,13,15-17]:

$$\Delta(\mathcal{A}\Psi) = 0 \quad \text{or} \quad \Delta \mathcal{A} = -\mathcal{A}_0 \frac{\Delta \Psi}{\Psi} , \qquad (23)$$

where $\mathcal{A}_0$ is a characteristic action value that may contain a numeric constant reflecting interaction details (thus, $\mathcal{A}_0 = ih/2\pi = i\hbar$ at the lowest, "quantum" complexity levels). Combining now eqs. (22) and (23), we obtain the "wavefunctional" counterpart of the universal Hamilton-Jacobi equation in the form of *universal Schrödinger equation* for the generalised wavefunction (or distribution function):

$$\mathcal{A}_0 \frac{\partial \Psi}{\partial t} = \hat{H}\left(x, \frac{\partial}{\partial x}, t\right) \Psi(x,t) , \qquad (24)$$

where continuous derivative notations are used and the Hamiltonian operator $\hat{H}$ is obtained from the Hamiltonian $H(x,p,t)$ by causal quantization. The generalised Schrödinger equation is completed by the *generalised Born rule*, obtained from the *dynamic matching conditions* for regular and intermediate realisations (they give the coefficients $c_i^r$ in the universal state-function expression, eq. (8)), and presenting the wavefunction, or its squared modulus, as *realisation probability distribution* [4-6,11,13,16,17]:

$$\alpha_r = |\Psi(X_r)|^2 , \qquad (25)$$

where $X_r$ is the *r*-th realisation configuration and one may have the value of the generalised distribution function itself at the right-hand side of eq. (25) for "particle-like" complexity levels.

Equations (21)-(25) constitute the basis of the *unified Hamilton-Schrödinger formalism* that should be accompanied by the unreduced, causally complete, and therefore inevitably dynamically multivalued, equation solution, such as the above result of the generalised EP method, eqs. (7)-(12). This universal formalism is a rigorous expression of the universal symmetry of complexity and unifies extended versions of various particular dynamic equations, usually corresponding to several first terms of power-series expansion of the generalised Hamiltonian [4-6].

Cosmological meaning of the universal symmetry of complexity goes, however, far beyond its particular mathematical expression. It represents the unified, exact *Order of the World*, applicable to the universe in the whole or any its part, including its *causally specified* origin and structure development (in their *realistic*, unreduced versions). Symmetry of complexity rigorously excludes, in particular, any possibility of universe emergence from "nothing" (with the zero total energy value), since only *positive* (and big) values of the initial interaction complexity (in the form of dynamic information) can give rise to further structure development (with equally *positive and big total energy*) and real time flow, eq. (22) (see also section 4). This fundamental positiveness of the universe content, distinguishing it from the zero-content unitary models, is directly related to the dynamic multivaluedness and intrinsic randomness of any real process, reduced to the dynamically single-valued projection in the unitary perturbative schemes that avoid any real, change-bringing interaction. We shall see below that the properties of the unreduced, dynamically multivalued world dynamics often do not even contain, or permit us to consistently solve, the accumulating "old" and "new" problems of the unitary cosmology and astrophysics, including the "missing" mass and energy content of the world.

## 3. Global properties of the emerging complex-dynamical universe structure

We shall outline, in this section, the "global", cosmological properties of the real, complex-dynamical world construction, such as they follow from the unreduced, multivalued dynamics of the underlying protofield interaction process (some of them were already mentioned in section 2). Note that practically none of this real-world properties can be consistently reproduced by any version of the unitary theory, irrespective of whether it is recognised as a true cosmological problem or not. Artificial addition of new abstract entities (such as "hidden dimensions" or new, equally "invisible" particle species), accompanied by "facilitated" parameter adjustment, certainly cannot change this situation, since new entities create new difficulties, thus simply displacing, or renaming, previous problems that remain basically unsolved because of their deceptive reduction to over-simplified, effectively zero-dimensional models.

### 3.1. Physically real, 3D space and irreversible time flow

We have seen in section 2 how the unreduced interaction between two initially homogeneous protofields gives rise to highly inhomogeneous structure of physically real, *tangible* space and equally real, but *immaterial*, *irreversibly flowing* time that can *not* be *really* "mixed" with space in an (abstract) "manifold".

We causally derive the *exact number (three) of spatial dimensions*, or "degrees of freedom", as being due to the dynamic entanglement of two protofields and their physically real interaction as such. This conservation of the number of basic entities, or "degrees of freedom" during the interaction process is a consequence of the universal symmetry (conservation) of complexity (see the end of section 2), supported by the *totality* of existing observations. We thus reveal also the detailed *physical nature* of those *emerging* space "dimensions" (remaining only abstract symbols in the canonical theory): they are obtained as a chaotically changing, interaction driven "mixture" (and inhomogeneous "deformation") of the physically real, initially homogeneous protofield "material".

We reveal the role of *essential nonlinearity*, omnipresent *dynamic instability*, and resulting *causal randomness (chaoticity)* of *quantum beat* dynamics of interacting protofields in establishment of *spatially chaotic* sequence of reduction-extension *events* within each field-particle, which gives rise to *unceasing* and *unpredictable* in detail (and *therefore irreversible*) *time flow*.

*Universality* of the obtained concept of space and time is supported by its unrestricted applicability to any system or level of complexity, giving rise to the fractally structured *hierarchy of space and time* that simply repeats the hierarchy of world (interaction) complexity and demonstrates the dynamic origin and close connection between space and time elements at each level of world structure. All cosmological problems of time (its absence in the effectively empty world, magically "tunneling" from nothing, etc.) are thus consistently solved (see the "time flow" condition of eq. (22) and section 4.1 for more details). Another aspect of time and space universality refers to the basic similarity of their properties (especially at most fundamental levels) throughout the whole vast, "physically infinite" universe. Postulated as "self-evident" in the canonical theory, this very *special* property finds now its substantiation in the *physically unified* structure of the underlying protofield system and related complex-dynamic *synchronisation of all individual quantum beat processes* (up to phase inversion), which determine the real time flow [4,12,13,17].

## 3.2. Unified dynamic origin of particles, interactions, and constants

It is important that the two omnipresent, "pervasive" *manifestations* of unreduced dynamic complexity, space and time, emerge in the protofield interaction process *in intrinsic unity* with the simplest *structures* of the first level of complexity, *elementary field-particles*, and their *fundamental properties* (mass, energy, motion, electric charge, spin, etc.), particle *interactions* with their observed properties (number, range, relative magnitude, unification), and *dynamical laws* (quantum and classical mechanics, special and general relativity), *all* of them being now *causally* and *explicitly obtained (derived)* from the fundamental interaction dynamics (*without any* "postulates") and thus naturally *unified* (section 2) [4,5,11-17]. The fundamental (measured) properties of real world structures are different, but related *measures* of the *same*, *universally* defined *dynamic complexity*, while structures themselves and their interactions represent two universal, dual *forms of complexity*, dynamic entropy and dynamic information, respectively, which are *permanently transformed* into one another according to the underlying *unique* "order of the world", the universal *symmetry of complexity*. Omitting detailed discussion of this *intrinsically unified* world structure and dynamics (section 2), we note only the *indispensable* role of omnipresent *dynamic multivaluedness* and the ensuing *chaoticity*, *diversity* (multiplicity) of forms and *adaptability* of real interaction products (absent in *any* unitary model), starting from the *quantum beat* process that constitutes the *causally complete* structure of (massive) elementary particles.

The related "difficult" problems of the unitary cosmology, which are *naturally resolved* in our complex-dynamical description, include the *problem of the universe wavefunction* and *quantization of gravity*. The universe wavefunction is *causally specified* now as the intermediate realisation of quantum beat processes in the *physically unified* protofield system. It naturally loses its *quantum* meaning there where classical (bound) systems start to emerge, but the *generalised* wavefunction and Schrödinger equation (see the end of section 2) re-emerge at each higher complexity level. As for the problem of *quantum* gravity, our universal gravitation *in general* is an indirect relation between *naturally discrete* quantum beat processes through the gravitational protofield and has therefore causal *quantum origin* from the beginning [4,5,12,13,17].

An essential novelty of the complex-dynamic cosmology is that it shows the *physical* origin of so-called *universal constants* and their *universality*, eventually reduced to the physically unified origin of the universe structure. We have seen above (section 2) that one of the constants, the *speed of light c*, is introduced in our theory not as an abstract, postulated "limit to signal speed" (standard relativity), but as a "normal", physical speed of signal propagation in the e/m protofield coupled to the gravitational medium, while time relativity and related limit to signal propagation velocity are consistently *derived* from the underlying *complex* interaction dynamics [4,5,12-17]. We can obtain the physical meaning of Planck's constant $h$ and fine structure constant $\alpha$, if we represent the well-known relation between $\hbar = h/2\pi$, $\alpha$, and elementary charge $e$ in a new form, involving electron rest mass $m_0$ and Compton wavelength $\lambda_C = h/m_0 c$:

$$\alpha \hbar = \frac{e^2}{c} \quad \Rightarrow \quad m_0 c^2 = \frac{2\pi}{\alpha} \frac{e^2}{\lambda_C} = N_\Re \frac{e^2}{\lambda_C} \;, \tag{26}$$

where $N_\Re = 2\pi/\alpha \approx 861$ emerges now as *realisation number of the electron* as a complex-dynamic process (of quantum beat), so that $\alpha$ coincides, up to a factor of $2\pi$, with the realisation probability $\alpha_r$ (see eq. (12a)), $\alpha = 2\pi\alpha_r$, while $\lambda_C$ is the length of a single "quantum jump" of the virtual soliton. We can write eq. (26) also as $\lambda_C = N_\Re r_e$ (where $r_e = e^2/m_0 c^2$ is the usual "classical radius" of the electron), which means that the size of the virtual soliton $D_e$ can be estimated as $D_e = 2\pi r_e = \pi d_e$. The meaning of Planck's constant follows from yet another form of eq. (26) and Compton wavelength expression:

$$h = \lambda_C p_0 = N_\Re \frac{e^2}{c} \;, \tag{27}$$

where $p_0 = m_0 c = E_0/c$. Planck's constant measures therefore the "volume" of the EP well for *any* field-particle, with the width of $\lambda_\text{C}$ (or $N_\Re$) and the depth of $p_0$ (or $e^2/c$). One can see the true origin of the extraordinary universality of $h$, remaining totally "mysterious" in the standard theory: the protofield deformation for various particles and processes occurs so that the EP well volume, $h$, remains the same (for the fixed protofield system parameters of a given world), whereas its depth (particle mass or charge) and width (number of realisations or length of virtual soliton jump) can vary considerably. This result is additionally confirmed by the ensuing consistent explanation of the *largest possible nuclear mass* as being roughly equal to that of the heaviest elementary particle (~ 100 GeV) [15,17].

Finally, the universal gravitational constant $\gamma$ of classical Newton's law of gravitation is used, together with $\hbar$ and $c$, in the canonical expressions for Planckian units, underlying many basic constructions of the scholar cosmology and particle theory and giving hugely exaggerated, too big or small units of length, time, and mass, separated by many orders of magnitude from any really observed (and all necessary) values (cf. e.g. the "hierarchy problem" in the particle mass spectrum). We can see now the origin of those difficulties and genuine involvement and meaning of gravity constant: whereas Planckian units describe *individual* EP well (quantum beat) dynamics *within each particle*, the usual gravitational constant corresponds to *indirect*, much weaker interaction between *different* particles (quantum beat processes) involving local protofield parameters and gravitational medium dynamics. Therefore one should use another, effective value of "gravitational constant", $\gamma_0$, in the Planckian unit combinations, which expresses the *direct*, much stronger *protofield attraction* as the *dynamically unified* origin of *all* particle interactions within the squeezed state of virtual soliton. It gives just the right values for Planckian units of length $L_\text{P}$, time $T_\text{P}$, and mass $M_\text{P}$, equal to extreme values of observed quantities $l_\text{exp}$, $t_\text{exp}$, and $m_\text{exp}$:

$$L_\text{P} = \left(\frac{\gamma_0 \hbar}{c^3}\right)^{\frac{1}{2}} \approx 10^{-17} - 10^{-16}\,\text{cm} \approx l_\text{exp} \ , \tag{28a}$$

$$T_\text{P} = \left(\frac{\gamma_0 \hbar}{c^5}\right)^{\frac{1}{2}} \approx 10^{-27} - 10^{-26}\,\text{s} \approx t_\text{exp} \ , \tag{28b}$$

$$M_\text{P} = \left(\frac{\hbar c}{\gamma_0}\right)^{\frac{1}{2}} \approx 10^{-22} - 10^{-21}\,g \ (10^2 - 10^3 \ \text{GeV}) \approx m_\text{exp} \ , \tag{28c}$$

where the relation between $\gamma_0$ and $\gamma$ can be specified, for example, using the values of ordinary Planckian unit of length $l_\text{P}$ and measured length $l_\text{exp}$: $\gamma_0 = (l_\text{exp}/l_\text{P})^2 \gamma \approx (10^{33} - 10^{34})\gamma$. The conventional difficulties are resolved thus without artificial introduction of any abstract entities (e.g. "hidden dimensions" in "brane-world" models), which create inevitably new difficulties and actually replace *dynamic* dimensions of the *multivalued* reality, incorrectly thrown off in its dynamically single-valued imitations. One can easily deduce major (fatal) consequences for the parts of the standard theory relying essentially upon the (usual) Planckian units, such as cosmological "inflation" and quantum gravity theories.

### 3.3. Self-tuning universe structure

It is evident already in terms of general logic that a *dynamically emerging* universe should have a dynamically consistent, *self-tuning*, adaptable structure, since this is the *essence* of genuine, *autonomous* structure formation as such. No wonder then that this is the property of complex-dynamic universe structure explicitly obtained in the protofield interaction process (section 2), as it is demonstrated by the dynamic origin of main entities, properties, and universal constants (section 3.2). Moreover, this *universal* property of the unreduced complex dynamics is preserved at any higher level of the emerging world structure. By contrast, it is *impossible* to obtain a feasible, stable universe structure in *any* unitary model, since its effectively zero-dimensional space does not leave any possibility for intrinsic adaptability. Mechanistic adjustment of artificially introduced entities and parameters can provide only a *basically inefficient* substitute for dynamical tuning, which gives the well-known "anthropic" difficulties of the unitary cosmology.

As can be seen from the self-consistent structure of the unreduced EP formalism (eqs. (7)-(12)), a viable universe with the same basic properties will always emerge for *generic* protofield interaction parameters.

According to the universal symmetry of complexity (section 2), greater quantities of dynamic information (generalised "potential energy") in the initial system configuration $V_{\text{init}}$ will lead to bigger dynamic entropy (generalised mass-energy) of the emerging universe structure $M_{\text{univ}}$:

$$V_{\text{init}} = M_{\text{univ}} c^2 \,, \tag{29a}$$

where the emerging structure quickly ramifies into the probabilistic (multivalued) fractal hierarchy of higher complexity levels, maintaining the same principle of intrinsic adaptability:

$$M_{\text{univ}} \to \sum_{\text{part}} N_{\text{part}} m_{\text{part}} + \frac{V_{\text{fund}}}{c^2} \to \sum_{\text{atom}} N_{\text{atom}} m_{\text{atom}} + \frac{V_{\text{chem}}}{c^2} \ldots \,, \tag{29b}$$

with "part" and "atom" designating progressively emerging species of elementary particles (and their interactions $V_{\text{fund}}$), atoms (and their interactions $V_{\text{chem}}$), and so on. Since both $V_{\text{fund}}$ and particle masses at the first complexity level depend (through the protofield tension) on the number of particles formed, the latter will be limited quantitatively and qualitatively (in the number of stable particle species). While quantitative aspect is more evident and corresponds to a general balance of eqs. (29), qualitative aspect provides a *causal* explanation of observed instability of all particle species but one shallow-EP (leptonic) species, known as the electron, and one deep-EP (hadronic) species, represented by proton.

Exceptions from generic rules can exist rather for extreme values of protofield interaction magnitude, but they also find their suitable places in the holistic complex-dynamical world picture. Ultimately strong protofield interaction will create a macroscopically large, "many-particle" protofield "collapse" that may have a number of different degrees [4], from a partially coherent "condensate" of elementary particles ("superdense" cosmic objects, such as "neuron stars"), which is still a part of "ordinary" reality, to the total protofield collapse down to their "pre-interaction" state of the unique "proto-matter", which does not contain anything from this world and should be considered as effective nothingness with respect to it. Contrary, to abstract and contradictory "exact solutions" of the unitary theory (such as "black holes"), each of these states can be provided with the causal, physically specified basis of its origin and structure, showing qualitative correlations with a number of observed "exotic" objects of the universe (e.g. quasars) and their features. The case of ultimately weak protofield interaction corresponds to small fluctuations of their structure that cannot transform into real, massive matter and may account for either "primordial" state of the protofields or, more realistically, observed universe state away from any massive matter, in the "vacuum", including propagating ordinary photons and in particular the so-called "microwave radiation background" related in the standard cosmology to the "remnants" of the first stages of the Big Bang and its properties.§ We see now that in the causally emerging, interaction-driven universe structure such "vacuum fluctuations" are inevitable and should not be directly related to a particular universe dynamics or evolution feature.

Note finally the huge, *exponentially large efficiency* of complex-dynamic adaptability (self-tuning) process: it is due to the unceasingly breeding and permanently changing realisations of the probabilistic dynamical fractal (section 2), which gives rise to the real-time, "fantastically efficient" exploration by the system of (almost) all existing possibilities for structure development [5].

## 4. Unified complex-dynamic solution to the problems of mass, energy, and entropy

### 4.1. Universe energy positivity and the time arrow

According to the universal symmetry (conservation) of complexity (see the end of section 2), the total dynamic complexity does not change in a structure emergence process, but is transformed instead from its "latent" (but real and *positively* defined) form of dynamic information (universally expressed by the generalised action quantity) into the "unfolded" form of dynamic entropy. Therefore any "compensation" of the positive total energy of moving bodies by the negative energy of their gravitational attraction, giving zero value of total energy, as it is implied by the unitary cosmology, is impossible in the real world dynamics. In fact, this "zero-energy solution" is due to the zero-complexity reduction of the dynamically single-valued model of the standard theory. By contrast, the inevitable positivity of the total complexity-energy of any real system is due to its dynamically multivalued, and therefore chaotic, dynamics, where the "thermal energy" of chaotic realisation change always determines the large positive balance of the total energy.

---

§Note that some versions of the unitary theory make reference to vacuum fluctuations of "zero-point field" or "space-time foam" as being due to formal solutions of eventually postulated equations. We should emphasize the totally causal origin of our weak interaction limit within the same interaction process between two protofields but at small values of effective coupling, for which any essentially nonlinear protofield deformation, and therefore quantum beat dynamics, is impossible.

This energy positivity condition is directly related to the direction of the arrow of time (and the very existence of time flow), by a rigorously derived and absolutely universal relation of eq. (22), which means that the positive stock of total energy-complexity gives rise to the flow of time as such, since for the system globally at rest, $\Delta t = -\Delta \mathcal{A}/E$ and with $\Delta \mathcal{A} < 0$ (because of dynamic multivalued-ness) $\Delta t > 0$ only if $E > 0$. In other words, a universe with zero total energy *could not exist* at all, in any configuration. Moreover, a small positive energy will give rise to proportionally small mass-energy content of the universe. This fundamentally substantiated conclusion about the *real, dynamically multivalued* universe emergence and structure puts an end to various formal postulates and hypotheses of unitary cosmology about possibility of universe appearance from nothing by a sort of "quantum tunneling", or "vacuum fluctuation", based on the assumption about zero total energy (where positive mass-energy of "matter" is compensated by the negative energy of gravitational attraction). Such is the well-known Wheeler-DeWitt equation and related ideas of the unitary "quantum cosmology". Even when a unitary theory inserts a positive energy in its formally postulated equations, it does not see the genuine physical origin and meaning of both energy/mass and its positivity, losing the main, chaotic part of system dynamics. Indeed, the assumed unitary "compensation" is impossible because the dynamically multivalued, chaotic part of any dynamics adds a dominating positive part to the total energy balance. We shall see that this loss of the main part of energy and motion in the unitary theory underlies *all* "difficult" problems of cosmology and astrophysics: mass and energy are lost in the unitary universe models *from the beginning*, and there is no wonder that various aspects of this basic deficiency emerge inevitably with growing precision and completeness of measurements.

Another aspect of the positive complexity-energy and time arrow of a real universe is a permanent, *strictly positive growth of dynamic entropy* accompanying *any* structure emergence, which resolves the old contradiction of the unitary science between the "second law" and apparently "growing order" during structure formation. Any unitary structure is basically regular only because of artificial limitation (dynamic single-valuedness) of the unitary theory itself, while the unreduced analysis of structure creation process shows (section 2) that any, even most externally regular structure, can appear and exist only due to the dominating *internal chaoticity* of its (very similar) realisation change (which is a limiting regime of "multivalued self-organisation") [4-7]. It is yet more important that this omnipresent entropy growth constitutes only a part of the whole *symmetry*, or conservation, of complexity (again contrary to the unitary science paradigm), since it occurs at the expense of equal decrease of the initial dynamic information of the system interaction configuration. The universe, its *real* structure, evolution, and any part dynamics are based, therefore, on the absolutely general and *exact* (never broken) principle of symmetry, the symmetry of (unreduced) complexity, constituting thus the genuine *Order of the World* that possesses the intrinsic, autonomous *structure creation power*.

### 4.2. Locally missing mass: unitary model deficiency

The so-called *dark mass problem* involves various observation data showing that *local* cosmic structure dynamics (mostly for galaxies) would need much larger (from several to hundreds times more) quantities of ordinary, massive matter, than those that can actually be viewed or deduced within any reasonable assumption (see e.g. refs. [25-27]). Big variation of the missing mass effect between various cosmic structures is another characteristic, and puzzling, feature of the problem. We can show that these difficulties of the unitary theory originate from the same its incorrect neglect of the main, chaotic part of system dynamics, now occurring at the level of a local cosmic object dynamics. If one considers the real, dynamically multivalued system behaviour, the problem will not appear and the truly chaotic dynamics of real objects will account for observed dynamical features with the "visible", normal mass values. It is important that one should take into account the *genuine, dynamically multivalued* chaos in a many-body system, rather than one of its unitary imitations by an "involved", but basically regular behaviour.

The main idea is physically straightforward: because of artificial cut of all system realisations but one in the unitary theory (this is an *exponentially big* reduction for a many-body system), one obtains inevitably a "missing motion" problem, which is *interpreted* as mysteriously "missing mass" within the same unitary imitation. One can specify this result in various ways, and we start with a demonstration of incompleteness of the standard "virial theorem" application to the real, multivalued dynamics of a many-body system, since it shows how the major "balance" between potential and kinetic energy can be modified by the *true* chaos.

If system components move under the influence of gravitational attraction, e.g. in a galaxy, then the ordinary virial theorem gives the following relation between the time-averaged values of kinetic $\bar{T}$ and potential $\bar{U}$ energy of a system or any its component (see e.g. [28]):

$$2\bar{T} = -\bar{U}, \tag{30}$$

whereas in reality this *regular* kinetic energy, $\bar{T} = \bar{T}_{\text{reg}}$, is a *small* part of its true, *chaotic* content $\bar{T}_{\text{real}}$:

$$\bar{T}_{\text{real}} = \bar{T}_{\text{reg}} N_\Re, \tag{31}$$

where $N_\Re$ is the effective number of system realisations for a given type of observation and corresponding "averaging" (usually $N_\Re \gg 1$, while $N_\Re = 1$ for the unitary approximation of the standard theory). The *observed* potential energy, $\bar{U}_{\text{obs}}$, gives *real* kinetic energy:

$$2\bar{T}_{\text{real}} = -\bar{U}_{\text{obs}}. \tag{32}$$

But when observations are interpreted within the unitary, deficient version of dynamics, eq. (30), stating that

$$2\bar{T}_{\text{reg}} = -\bar{U}_{\text{obs}}, \tag{33}$$

one obtains a discrepancy, $\delta$, dividing eq. (32) by eq. (33):

$$\delta = \frac{\bar{T}_{\text{real}}}{\bar{T}_{\text{reg}}} = N_\Re. \tag{34}$$

It is explained *within* the unitary model as being due to "invisible", but actually present, or "dark" mass, $M_{\text{dark}} = M_{\text{real}} - M_{\text{reg}}$, whose relative value can be estimated as

$$\frac{M_{\text{real}}}{M_{\text{reg}}} = \frac{\bar{T}_{\text{real}}}{\bar{T}_{\text{reg}}} = \delta = N_\Re. \tag{35}$$

Therefore, the observed discrepancy can be used, within the *complex-dynamic interpretation*, for estimation of effective $N_\Re$ values. Since $\bar{T} \propto \overline{Mv^2}$, one can say that in reality there is *too much motion*, or (deviating) *velocity*, in a system with respect to unitary expectations, so that one has rather a "dark velocity" effect:

$$\left(\overline{v^2}\right)_{\text{real}} = N_\Re \left(\overline{v^2}\right)_{\text{reg}}. \tag{36}$$

One can easily refine this result for a distance-dependent case, $N_\Re = N_\Re(r)$ (where $r$ is a coordinate within the system), involving the popular formulation in terms of velocity-distance dependence curves, or "rotation curves", for galaxies. In that case an "anomalous" $v(r)$ dependence is not due to anomalies of mass distribution, $M(r)$ (attributed to "dark matter halos"), but due to "unexpected" (in the unitary model) contribution to *average* velocity from chaotic motion parts, so that $v(r)$ is proportional not to $\sqrt{M_{\text{reg}}(r) + M_{\text{dark}}(r)}$, but to $\sqrt{N_\Re(r)}$. In a general case,

$$v(r) = \sqrt{\frac{\gamma N_\Re(r) M_{\text{obs}}(r)}{r}} \quad \text{or} \quad N_\Re(r) = \frac{r v^2(r)}{\gamma M_{\text{obs}}(r)}, \tag{37}$$

where $M_{\text{obs}}(r) = M_{\text{real}}(r)$ is the *ordinary*, "visible" mass within radius $r$, and one can *derive* the features of *chaotic* system dynamics, $N_\Re(r)$, from the observed $v(r)$ and $M_{\text{obs}}(r)$ dependences for perceivable, "normal" object components. As should be expected, $N_\Re(r)$, and thus chaoticity, will typically have a wide, sometimes irregular maximum in galactic halos and central, inter-component regions of a cluster.

The observed big variations of "dark mass" effects for different objects represent a "heavy" difficulty for any explanation in terms of additional, "invisible" entities, but are, on the contrary, *inevitable* for the above *unified* explanation in terms of the true (multivalued) chaos effects that *should* vary a lot. Moreover, one can trace a definite qualitative correlation between the expected object chaoticity (degree of irregularity), its spatial dependence, and the observed magnitude of "missing mass" effects (extended verification is certainly necessary). One may note also that it is much more consistent to explain an observed, *variable* system *property* by a fundamental *property* of its dynamics, rather than by a new, strangely escaping, and inevitably *fixed entity* (this refers also to interpretation of the very *origin of mass* in the universal science of complexity and unitary field theory [4,5,12-17]). One should also take into account the spatial dependence of chaotic mass distribution effects (or "structural" chaos) that tend to accumulate just outside of the main mass and interaction concentration in the system (especially for those with a "centred rotation" configuration), in agreement with data interpretation using eqs. (37). Note finally the deep *conceptual relation* between the missing mass effects at different levels of world dynamics, the missing (total) mass-energy of the universe (section 4.1), missing *dynamic* origin of particle mass (replaced by the artificially introduced *new entity* of "Higgs boson"), and the "dark mass" effects at the level of cosmic objects, obtained in the universal science of complexity by the *unified* and consistent explanation of *all* those properties in terms of *multivalued*, chaotic dynamics of *any* real system, *rigorously derived* by the *unreduced* problem solution (cf. section 1).

## 4.3. Globally missing energy and Big Bang contradictions: deficient linearity

The origin of the *globally* missing, "distributed" energy, or "dark energy" [25-27], that could also be called "missing universe acceleration", is directly related to the vicious circle of the unitary cosmology scheme centred on the Big Bang hypothesis, or "exploding vacuum" solution. Indeed, the latter starts from *postulated*, artificially imposed *nothingness of the essential universe content* (section 4.1), in the form of dynamically single-valued, zero-complexity reduction of universe dynamics (irrespective of particular "model" details). Because of the intrinsic instability of that fundamentally fixed, static construction, one is obliged to further impose a mechanistic "general expansion" (or the reverse squeeze) of the universe as a single possible mode of its (totally illusive) "development". The choice for expansion, or Big Bang, is justified by a *particular interpretation* of the observed "red shift" effect (the interpretation that involves a number of serious contradictions in itself). However, the intrinsic *conceptual instability* of *any* unitary model (absence of adaptable *degrees of freedom*, rather than abstract "parameters") persists in the form of multiple particular problems of the Big Bang model whose proposed "solutions" only transfer the difficulties to other formulations or artificially introduced entities. The "dark energy" problem represents only the latest in the list, though scandalously big and long hidden, rupture in the basically frustrated construction: a *slightly* uneven red-shift dependence on distance leads to a *huge deficiency* in the source of corresponding uneven expansion, supposed to be a distributed stock of mysterious, *invisible* energy that should inevitably take *very exotic*, normally *impossible* forms. This final, and apparently "definite", impasse of missing energy (and mass) content of the universe simply takes us back to the beginning of the unitary vicious circle, where such emptiness of the universe content has been *explicitly inserted* by the unitary paradigm (this is but a degenerate case of the complexity conservation law, astonishing in its long-lasting simplification, $0=0$).

By contrast, the unreduced, dynamically multivalued and probabilistically fractal structure of real interaction dynamics leads to *globally stable* concept of universe structure development, just because it is based on the *omnipresent* and massively adaptable *local*, dynamic instability. The explicit universe structure emergence in the initially homogeneous system of interacting protofields, starting from the physically real space, time, and elementary particles, intrinsically unified with their fundamental properties and interactions, can be described as a distributed *im*plosion of ubiquitous structure creation, as opposed to mechanistic, and intrinsically *destructive*, rather than creative, *ex*plosion of the Big Bang (and "inflation") schemes.

Therefore the "dark energy" problem *does not even appear* in the complex-dynamic, intrinsically creative cosmology. The self-tuning universe structure, liberated from unitary instabilities and related "anthropic" speculations, emerges naturally and self-consistently, simply due to the unreduced, truly "exact" picture of the underlying generic interaction. As for the origin of the observed red shift effect in *radiation* spectra of *distant* objects, it finds its consistent explanation in terms of *nonlinear radiation propagation* properties in the system of coupled protofields, where some (relatively weak) loss of energy by soliton-like photons, propagating in the e/m protofield medium, is *inevitable* because of their *irreducible*, though relatively small, coupling to the degrees of freedom of the gravitational medium. Note the essential difference of this *nonlinear* energy dissipation mechanism and result from linear scattering effects in any ordinary, "corpuscular" model. The soliton-like photon, remaining *stabilised* by interaction with the gravitational protofield, can slowly give its energy to the gravitational degrees of freedom *without* any noticeable change of its direction of propagation (i.e. without any resulting "blur" effects in the distant object images). Characteristic "transpiercing" and "circumventing" modes of soliton interaction with "weak" enough obstacles can explain anomalously small loss and vanishing angular deviation effects for photons and very high-energy particles (see below). Detailed calculations of the effect will inevitably involve many unknown parameters of the system, but qualitative properties and consistency of the whole picture provide convincing evidence in favour of this *fundamentally new* explanation of the red shift effect and its further refinement.

In particular, the *nonlinear* red shift dependence on distance that gives rise to *catastrophic* consequences in the unitary cosmology can only be *natural* in the complex-dynamical, *essentially nonlinear* picture (section 2). The nonlinear energy-loss mechanism of soliton-like photons explains why this loss grows *more slowly* with distance, than any usual mechanism of diffuse scattering would imply (cf. the above note on soliton scattering dynamics). Similar dynamics could solve, by the way, the persisting puzzle of GZK effect for the ultra-relativistic particles, since at those super-high energies the motion of a massive particle approaches that of (a group of) photons, according to the results of quantum field mechanics [4,12-17]. Another, though maybe less important, feature of red-shift data correlating with our explanation is the apparent *growth of average scatter* of data points with distance.

## 5. Conclusion

Returning to the general picture of our emerging universe, note once more that it does not contain "motion-on-circles" dynamics, on any scale of structure creation, so that the initial amount of dynamic information, in the form of protofield interaction, gives rise to generalised, complex-dynamical system *birth*, followed by its gradual, *irreversible*, and "global" transformation into dynamic entropy (developed structure) representing a universally defined, *finite* system *life*, which ends up in the state of generalised *death*, or *equilibrium*, around the total transformation of the initial dynamic information into entropy (unless additional dynamic information is introduced into the system) [4]. The generalised "potential energy" of interacting protofields can be introduced e.g. by their explicit separation from the "pre-existing" state of "totally unified" (mixed) protofields that could have the form of a generally inert quark-gluon "condensate" in its "absolute" ground state (but these "prehistoric" assumptions are subject to inevitable uncertainty and can be estimated rather by general consistency and parsimony principles). What appears to be much more certain, however, is that one *does* need an initial form of "potential" interaction energy, *positively* defined and universally specified here as "dynamic information", since the birth of a structured, real universe from essential "nothingness", without *genuine* interaction development (which is the preferred dogma of the conventional unitarity), contradicts the fundamentally substantiated and *universally* confirmed symmetry of complexity.

Finally, we may summarise additional empirical evidence in favour of our complex-dynamical universe description, whose consistent interpretation within the standard, unitary cosmology paradigm seems much less probable. The highly uneven, long-distance concentration of anomalous, super-intense "quasar" sources of energy (as well as their "peculiar" red-shift tendency) points to a (probably moving) "shape of the world", which looks quite natural in our interacting protofield logic, while it would need additional, "unnatural" assumptions in the Big Bang logic of "exploding emptiness". Growing problems with the age of the universe and its separate components can be naturally solved in our complex-dynamic cosmology, while the unitary theory seems to encounter here another series of its inbred "instabilities" (due to the rigidly fixed, mechanistic data fit). The same refers to structural difficulties of the omnipresent expansion and natural elimination in our approach of this and other "old" difficulties of the unitary theory, such as the average space flatness and homogeneity (some others are mentioned above and all are well-known). Intrinsic unification with realistic, universal solution of the stagnating problems of quantum mechanics, field theory, and relativity (sections 2, 3) constitutes the *unique* feature of our theory that, being certainly desirable, cannot be even expected for any existing or future version of the unitary imitation of world dynamics.

## References

[1] M. López-Corredoira, "Observational cosmology: caveats and open questions in the standard model", astro-ph/0310214 at http://arXiv.org.

[2] H. Bergson, *L'évolution créatrice*, Félix Alcan, Paris (1907). English translation: *Creative Evolution*, Macmillan, London (1911).

[3] I. Prigogine and I. Stengers, *Order Out of Chaos*, Heinemann, London (1984).

[4] A.P. Kirilyuk, *Universal Concept of Complexity by the Dynamic Redundance Paradigm: Causal Randomness, Complete Wave Mechanics, and the Ultimate Unification of Knowledge*, Naukova Dumka, Kiev (1997) 550 pp., in English. For a non-technical review see physics/9806002 at http://arXiv.org.

[5] A.P. Kirilyuk, "Dynamically Multivalued, Not Unitary or Stochastic, Operation of Real Quantum, Classical and Hybrid Micro-Machines", physics/0211071 at http://arXiv.org.

[6] A.P. Kirilyuk, "Universal symmetry of complexity and its manifestations at different levels of world dynamics", *Proceedings of Institute of Mathematics of NAS of Ukraine* **50** (2004) 821–828; physics/0404006 at http://arXiv.org.

[7] A.P. Kirilyuk, "Dynamically multivalued self-organisation and probabilistic structure formation processes", *Solid State Phenomena* **97–98** (2004) 21–26; physics/0405063 at http://arXiv.org.

[8] A.P. Kirilyuk, "The universal dynamic complexity as extended dynamic fractality: Causally complete understanding of living systems emergence and operation", In: G.A. Losa, D. Merlini, T.F. Nonnenmacher, and E.R. Weibel (eds.), *Fractals in Biology and Medicine. Vol. III*, Birkhäuser, Basel (2002) 271-284; physics/0305119 at http://arXiv.org.

[9] A.P. Kirilyuk, "Quantum chaos and fundamental multivaluedness of dynamical functions", *Annales de la Fondation Louis de Broglie* **21** (1996) 455-480; quant-ph/9511034 - 36 at http://arXiv.org.

[10] A.P. Kirilyuk, "Causal Wave Mechanics and the Advent of Complexity. IV. Dynamical origin of quantum indeterminacy and wave reduction", quant-ph/9511037 at http://arXiv.org.

[11] A.P. Kirilyuk, "Double Solution with Chaos: Dynamic Redundance and Causal Wave-Particle Duality", quant-ph/9902015.

[12] A.P. Kirilyuk, "Double Solution with Chaos: Completion of de Broglie's Nonlinear Wave Mechanics and Its Intrinsic Unification with the Causally Extended Relativity", quant-ph/9902016 at http://arXiv.org.

[13] A.P. Kirilyuk, "Universal gravitation as a complex-dynamical process, renormalised Planckian units, and the spectrum of elementary particles", gr-qc/9906077 at http://arXiv.org.

[14] A.P. Kirilyuk, "75 Years of Matter Wave: Louis de Broglie and Renaissance of the Causally Complete Knowledge", quant-ph/9911107 at http://arXiv.org.

[15] A.P. Kirilyuk, "100 Years of Quanta: Complex-Dynamical Origin of Planck's Constant and Causally Complete Extension of Quantum Mechanics", quant-ph/0012069 at http://arXiv.org.

[16] A.P. Kirilyuk, "75 Years of the Wavefunction: Complex-Dynamical Extension of the Original Wave Realism and the Universal Schrödinger Equation", quant-ph/0101129 at http://arXiv.org.

[17] A.P. Kirilyuk, "Quantum Field Mechanics: Complex-Dynamical Completion of Fundamental Physics and Its Experimental Implications", physics/0401164 at http://arXiv.org.

[18] L. Randall and R. Sundrum, "A Large Mass Hierarchy from a Small Extra Dimension", *Phys. Rev. Lett.* **83** (1999) 3370-3373; hep-ph/9905221 at http://arXiv.org.

[19] L. Randall and R. Sundrum, "An Alternative to Compactification", *Phys. Rev. Lett.* **83** (1999) 4690-4693; hep-th/9906064.

[20] V. Sahni and Yu. Shtanov, "New Vistas in Braneworld Cosmology", *Int. J. Mod. Phys.* **D11** (2000) 1515-1521; gr-qc/0205111 at http://arXiv.org.

[21] P.H. Dederichs, "Dynamical diffraction theory by optical potential methods", In: H. Ehrenreich, F. Seitz, and D. Turnbull (eds.), *Solid state physics, Vol. 27*, Academic Press, New York (1972) 136–237.

[22] A.P. Kirilyuk, "Theory of charged particle scattering in crystals by the generalised optical potential method", *Nucl. Instr. Meth.* **B69** (1992) 200-231.

[23] L. de Broglie, "Recherches sur la théorie des quanta", Thèse de doctorat soutenue à Paris le 25 novembre 1924, *Annales de Physique* (10e série) **III** (1925) 22-128. Reprinted edition: L. de Broglie, *Recherches sur la théorie des quanta*, Fondation Louis de Broglie, Paris (1992).

[24] L. de Broglie, "Waves and quanta", *Nature* **112** (1923) 540.

[25] S. Khalil and C. Munoz, "The Enigma of the Dark Matter", *Contemp. Phys.* **43** (2002) 51-62; hep-ph/0110122.

[26] K.A. Olive, "TASI Lectures on Dark Matter", astro-ph/0301505 at http://arXiv.org.

[27] V. Sahni, "Dark Matter and Dark Energy", astro-ph/0403324 at http://arXiv.org.

[28] L.D. Landau and E.M. Lifshitz, *Mechanics*, Nauka, Moscow (1988), fourth Russian edition.